\documentclass{article}      % Specifies the document class \usepackage{epsfig}
\pdfoutput=1
\usepackage{amsmath}

\usepackage{graphicx}
\usepackage{mathtools}

\title{q-Viscous Burgers' Equation: Dynamical Symmetry, Shock Solitons and q-Semiclassical Expansion}

\author{ Sengul Nalci Tumer and Oktay Pashaev \\Department of Mathematics, Izmir Institute of Technology \\ Urla-Izmir, 35430, Turkey}

\begin{document}
\newcommand{\be}{\begin{equation}}
\newcommand{\ee}{\end{equation}}
\newcommand{\bea}{\begin{eqnarray}}
\newcommand{\eea}{\end{eqnarray}}
\newcommand{\disp}{\displaystyle}
\newcommand{\la}{\langle}
\newcommand{\ra}{\rangle}

\newtheorem{thm}{Theorem}[subsection]
\newtheorem{cor}[thm]{Corollary}
\newtheorem{lem}[thm]{Lemma}
\newtheorem{prop}[thm]{Proposition}
\newtheorem{definition}[thm]{Definition}
\newtheorem{rem}[thm]{Remark}
\newtheorem{prf}[thm]{Proof}

\maketitle

%%%%%%%%%%%%%%%%%%%%%%%%%%%%%%%%%%%%%%%%%%%%%%%%%%%%%%%%%%%%%% % You may repeat \author \address as often as necessary      % %%%%%%%%%%%%%%%%%%%%%%%%%%%%%%%%%%%%%%%%%%%%%%%%%%%%%%%%%%%%%%

\begin{abstract}
We propose new type of $q$-diffusive heat equation with nonsymmetric $q$-extension of the diffusion term. Written in relative gradient variables this system appears as the $q$- viscous Burgers' equation. Exact solutions of this equation in polynomial form as generalized Kampe de Feriet polynomials, corresponding dynamical symmetry and description in terms of Bell polynomials are derived. We found the generating function for these polynomials by application of dynamical symmetry and the Zassenhaus formula. We have constructed and analyzed shock solitons and their interactions with different $q$. We obtain modification of the soliton relative speeds depending on value of $q$.For $q< 1$  the soliton speed becomes bounded from above and as a result in addition to usual Burgers soliton process of fusion, we found a new phenomena, when soliton with higher amplitude but smaller velocity is fissing to two solitons. q-Semiclassical expansion of these equations are found in terms of Bernoulli polynomials in power of $\ln q$.
\end{abstract}

\section{Introduction}

The heat equation and its modifications are the simplest equations in mathematical physics, modelling diffusion heat transfer and other phenomena. To model more reach class of diffusion phenomena, several extensions of the diffusion equation by fractional calculus, quantum or $q$-calculus, noncommutative calculus, etc. were proposed \cite{Nalci Pashaev}, \cite{Pashaev Nalci},\cite{Martina Pashaev}. Described in terms of relative gradients, the heat equation appears in the form of nonlinear Burgers' equation. Solution of this Burgers' equation as the  shock solitons and their interactions play fundamental role in description of soliton phenomena. Extensions of this equation by the $q$-deformations lead to a new type of soliton phenomena, like $q$-shock solitons \cite{Nalci Pashaev}, \cite{Pashaev Nalci}, noncommutative shock solitons \cite{Martina Pashaev}, etc. This is why, any exactly solvable extension of heat and Burgers' equations play essential role in description of new type of soliton interactions, exact solvability of corresponding equations and in modelling new physical phenomena associated with them.

Recently, several extensions of diffusion equation by the $q$-deformation of partial derivatives were proposed and exact solutions in the form of $q$-shock solitons were constructed and  represented in terms of $q$-special functions. By such an approach the $q$-deformation of classical damped oscillator  as the $q$-deformed oscillator was studied in \cite{Nalci Pashaev1}. The quantum versions of $q$-oscillator  have attracted attentions due to  relations with quantum groups and exact solvability for different realizations of quantum symmetry, as symmetrical \cite{Biedenharn},\cite{Macfarlane}, non symmetrical \cite{Arikcoon}, Fibonacci\cite{Arik1} and Golden calculus \cite{PashaevNalci}, etc. In the set of papers  by Man'ko  and coauthors \cite{Manko} a physical approach to $q$-oscillator as nonlinear oscillator was proposed.  It was shown in \cite{Pashaev1} that every integrable system in action-angle variables is described as a set of nonlinear oscillators and appears in the form of the $q$- or more generally, the $f$- oscillator. In  paper \cite{Pashaev1},\cite{Pashaev3} the linear Schr\"{o}dinger equation with $q$-modified dispersion was proposed and the Madellung form of this equation as $q$-dispersive complex nonlinear Burgers' equation was derived.

In the present paper, following similar ideas we propose new type of heat equation with modified non-symmetric $q$-diffusive term.  This equation belongs to the heat hierarchy of infinite order diffusive equations.  Description of this equation in terms of relative gradients appears as a $q$-viscous Burgers' equation, which is a specific member of Burgers' hierarchy.  We are studying several classes of exact solutions, polynomial and shock soliton type. The polynomial solutions are generalizations of the Kampe de Feriet polynomials written in terms of Bell polynomials. We derive generating function for these polynomials by using the dynamical symmetry and the Zassenhaus formula. Related with this the generalized Boost operator, which generates exact solutions and the dynamical symmetry is explicitly constructed. Then we find one and two shock soliton solutions and study their interactions. We show that $q$-deformation modifies the speed of our solitons. Finally,  we develop the q-semiclassical expansion of our equations in $\lambda=\ln q$ as higher order deformations, written in terms of Bernoulli polynomials.

\section{q-Diffusive Heat Equation}

We introduce q-diffusive deformation of the heat equation in the following form
\begin{eqnarray}
\frac{\partial}{\partial t} \phi(x,t)=\left[\nu \frac{\partial^2}{\partial x^2}\right]_q \phi(x,t), \label{qdeformedheatdeformation}
\end{eqnarray}
where $\nu$ is diffusion constant and the $q$-operator
\begin{eqnarray}
\left[\nu \frac{\partial  ^2}{\partial x^2} \right]_q= \frac{q^{\nu \frac{\partial ^2}{ \partial x^2}}-1}{q-1} \label{qoperator}
\end{eqnarray}
is defined as a formal power series. In the limiting case $q \rightarrow 1$, equation (\ref{qdeformedheatdeformation}) reduces to the standard heat equation.

By the method of separation of variables we search solution of this equation in the form
\begin{eqnarray} \phi(x,t)= X(x) T(t).\nonumber \end{eqnarray}
Substituting this into  (\ref{qdeformedheatdeformation}) we get
\begin{eqnarray}
\frac{T^{'}(t)}{T(t)}= \frac{\left[\nu \frac{\partial^2}{\partial x^2}\right]_q X(x)}{X(x)}=-\lambda. \nonumber
\end{eqnarray}
As a result, we obtain two ordinary differential equations

\begin{eqnarray} & & T^{'}(t)+\lambda T(t)=0, \\
& & \left[\nu \frac{\partial^2}{\partial x^2}\right]_q X(x)+\lambda X(x)=0.
\end{eqnarray}
Solution of the first equation in $t$ is
 $$T(t)=e^{-\lambda t} T(0),$$
 where $T(0)$ is a constant.

\subsection{Finite Interval Case}
For the space part we consider the following eigenvalue problem on finite interval with the Dirichlet boundary conditions

\begin{eqnarray}  \left[\nu \frac{\partial^2}{\partial x^2}\right]_q X(x)&=&-\lambda X(x), \label{qspacepart} \\
X(0)=X(l)&=&0 .
\end{eqnarray}
In order to solve this problem we use the following boundary value problem
\begin{eqnarray}
& & -X^{''}(x)=\mu X(x), \nonumber \\
& & X(0)=X(l)=0  \label{standardspace},
\end{eqnarray}
with eigenvalues
\begin{eqnarray} \mu_n=\left(\frac{n \pi}{l}\right)^2,
\end{eqnarray}
and the corresponding eigenfunctions

\begin{eqnarray}
X_n(x)= \sqrt{\frac{2}{l}} \sin{\frac{n \pi}{l}x}.
\end{eqnarray}
This set of eigenfunctions is orthonormal and complete.
Then, substituting to equation (\ref{qspacepart}), and by using definition of the $q$-operator we obtain

\begin{eqnarray}
\left[\nu \frac{\partial^2}{\partial x^2}\right]_q X(x)&=&  \frac{q^{\nu \frac{\partial^2}{\partial x^2}} X(x)-X(x)}{q-1} \nonumber \\
&=& \frac{q^{-\mu \nu}-1}{q-1}X(x)= \left[ -\mu \nu \right]_q X(x),
\end{eqnarray}
which gives the relation between the eigenvalues of $q$-equation (\ref{qspacepart}) and equation (\ref{standardspace})
\begin{eqnarray}
\lambda = - \left[ -\mu \nu \right]_q.
\end{eqnarray}
As a result, solution of the $q$-deformed initial value problem (\ref{qspacepart})
 is obtained in terms of solution of standard Sturm-Liouville problem (\ref{standardspace}) with eigenvalues as $q$-numbers
$$\lambda_n= -[-\mu_n \nu]_q= - \left[-\left(\frac{n \pi}{l}\right)^2\nu\right]_q,$$ and the corresponding eigenfunctions
 \begin{eqnarray}
X_n(x)=\sqrt{\frac{2}{l}} \sin{\frac{n \pi}{l}x}.\label{eigenfunction}\end{eqnarray}
%(The solution is unique. Despite that (\ref{qspacepart}) is infinite order problem. This is easy to show by expanding solution in terms of orthonormal set of (\ref{eigenfunction}).)
It gives us particular solution of $q$-diffusive heat equation (\ref{qdeformedheatdeformation}) in the form
\begin{eqnarray} \phi_n(x,t)= X_n(x) T_n(t)= e^{-\lambda_n t} T(0)\,\, \sqrt{\frac{2}{l}}\,\, \sin{\frac{n \pi}{l}x},\nonumber \end{eqnarray}
where $\lambda_n= -[-(\frac{n \pi}{l})^2 \nu]_q.$

\begin{figure}[!tbp]
  \centering
  \begin{minipage}[b]{0.4\textwidth}
    \includegraphics[width=\textwidth]{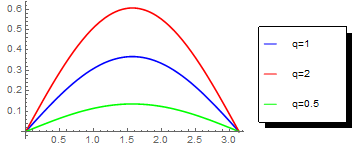}
    \caption{\small{Evolution of $n=1$ solution at time $t=1$}}
    \label{mode1}
  \end{minipage}
  \hfill
  \begin{minipage}[b]{0.4\textwidth}
    \includegraphics[width=\textwidth]{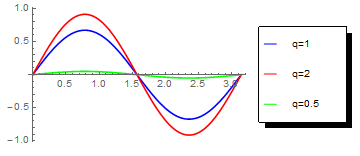}
    \caption{\small{Evolution of $n=2$ solution at time $t=0.1$}}
    \label{mode2}
  \end{minipage}
\end{figure}
In Figures (\ref{mode1}) and (\ref{mode2}) we show particular solutions for $n=1$ and $n=2$ modes correspondingly, in $q>1,$ $q<1$ and $q=1$ cases. As we can see,
comparing with the usual heat equation with $q=1$,
depending on $q$ the decaying process is going faster for $q<1,$ or slower for $q>1.$

The general solution is a superposition of these solutions
\begin{eqnarray}
\phi(x,t)= \sum_{n=1}^\infty A_n e^{-\lambda_n t} \sin{\frac{n \pi}{l}x} = \sum_{n=1}^\infty A_n e^{t{[-(\frac{n \pi}{l})^2 \nu]_q} } \sin{\frac{n \pi}{l}x}.  \label{solution}
\end{eqnarray}
To fix the Fourier coefficients $A_n$ we pose the following IVP
$$\phi(x,0)= f(x),$$
so that we get
$$\phi(x,0)= f(x)= \sum_{n=0}^{\infty } A_n \sin {\frac{n \pi}{l}x}.$$
Then the coefficients are found as
$$A_m= \frac{2}{l} \int_0^{l} f(x) \sin ({\frac{m \pi}{l}x})\, dx ,$$
and  solution is obtained in the form

\begin{eqnarray}
\phi(x,t)=  \frac{2}{l} \sum_{n=0}^\infty \int_0^{l} dy\,  f(y) \sin ({\frac{n \pi}{l}x}) \,  \sin ({\frac{n \pi}{l}y}) \, e^{-\lambda_n t} .\end{eqnarray}

We define the Green function for equation (\ref{qdeformedheatdeformation}) as
\begin{eqnarray}
G(x,y;t)= \frac{2}{l} \sum_{n=0}^{\infty} \sin ({\frac{n \pi}{l}x})\,\, \sin ({\frac{n \pi}{l}y})\,\, e^{t{[-(\frac{n \pi}{l})^2 \nu]_q} },\label{green}
\end{eqnarray}
so that solution of IBVP is
\begin{eqnarray}
\phi(x,t)= \int_0^{l} G(x,y;t) f(y) dy.
\end{eqnarray}

The Green function (\ref{green}) as evident, satisfies $G(x,y;t)= G(y,x;t)$ and at initial time $t=0$ it is just the Dirac delta function
\begin{eqnarray}
G(x,y;0)=\frac{2}{l} \sum_{n=0}^\infty \sin(\frac{n \pi}{l}x) \sin(\frac{n \pi}{l}y)= \delta (x-y).
\end{eqnarray}

%\begin{figure}[!tbp]
%  \centering
%  \begin{minipage}[b]{0.45\textwidth}
%    \includegraphics[width=\textwidth]{DiracDelta}
%    \caption{\small{Dirac Delta Function for $q=1$}}
%    \label{delta}
%  \end{minipage}
%  \hfill
%  \begin{minipage}[b]{0.45\textwidth}
%    \includegraphics[width=\textwidth]{DiracDeltaq}
%    \caption{\small{Spreading of Dirac Delta function for $q$}}
%    \label{spreading}
%  \end{minipage}
%\end{figure}

%From Figure (\ref{spreading}) we see that it describes a spreading of original delta function, which comparing with the $q=1$ case (see Figure (\ref{delta})), is going faster for $q<1$ and slower for $q>1$ .

Due to relation
\begin{eqnarray} F\left(\frac{d}{d x}\right)e^{ikx}= F(ik)e^{ikx},\label{identity}\end{eqnarray}
which implies
 $$e^{t[\nu \frac{d^2}{dx^2}]_q} e^{\pm i \frac{n \pi}{l}x }= e^{t[\nu (\frac{i n \pi }{l})^2]_q} e^{\pm i \frac{n \pi}{l}x },$$
we can rewrite (\ref{green}) in an operator form by using the evolution operator

\begin{eqnarray}
G(x-y\,;t)&=& e^{t[\nu \frac{d^2}{dx^2}]_q} \delta(x-y)= \frac{2}{l} \sum_{n=0}^\infty  e^{t[\nu \frac{d^2}{dx^2}]_q} \sin(\frac{n \pi}{l}x) \sin(\frac{n \pi}{l}y) \nonumber \\
&=& \frac{2}{l} \sum_{n=0}^\infty e^{t[-\nu(\frac{n \pi}{l})^2]_q} \sin(\frac{n \pi }{l}x) \sin(\frac{n \pi}{l}y).
\end{eqnarray}

\subsection{Infinite Interval Case}
Now we consider the initial value problem for $q$-diffusive heat equation in infinite interval:

\begin{eqnarray}  \frac{\partial}{\partial t} \phi(x,t)&=&\left[\nu \frac{\partial^2}{\partial x^2}\right]_q \phi(x,t) , \nonumber \\
\phi (x,0)&=& f(x),\label{infiniteinterval} \\
-\infty<&x&<\infty.  \nonumber
\end{eqnarray}

By using the Fourier transform
\begin{eqnarray}\phi (x,t)= \frac{1}{\sqrt{2 \pi}} \int_{-\infty}^{\infty} dk\,\, e^{ikx}\,\, \tilde{\phi}(k,t) \label{fourier}\end{eqnarray}
and substituting into  (\ref{infiniteinterval})
we have
\begin{eqnarray}\int_{-\infty}^{\infty} dk\,\, e^{ikx}\,\, \tilde{\phi}_t (k,t)=\int_{-\infty}^{\infty} dk\,\, \tilde{\phi}(k,t) {\left[\nu \frac{\partial^2}{\partial x^2}\right]_q}\,\, e^{ikx}. \label{fourierpart1}\end{eqnarray}

Due to property (\ref{identity})

%\begin{prop}
%For any analytic function $F(x)$ we have
%\begin{equation}
%F(\frac{d}{d x})e^{ikx}= F(ik)e^{ikx}. \label{proposition}
%\end{equation}
%\end{prop}

%\begin{prf}
%If
%$F(x)$ is any function expandable to power series
%$ F(x)= \sum_{n=0}^{\infty} a_n x^n,$
%then we get the desired result as
%$$F(\frac{d}{dx}) e^{ikx}= \sum_{n=0}^{\infty} a_n \frac{d^n}{dx^n} e^{ikx}=\sum_{n=0}^{\infty} a_n (i k)^n e^{ikx} = F(i k )e^{ikx}.$$
% \end{prf}

\begin{eqnarray}\left[ \nu \frac{d^2}{d x^2}\right]_q e^{ikx}= \left[ \nu (ik)^2 \right]_q e^{ikx}= \left[ -\nu k^2\right]_q e^{ikx} \label{property1}.\end{eqnarray}
the integral (\ref{fourierpart1}) becomes

\begin{eqnarray}
\int_{-\infty}^{\infty} dk\,\, e^{ikx}\,\, \tilde{\phi}_t(k,t) = \int_{-\infty}^{\infty} dk \tilde{\phi} (k,t) \left[ -\nu k^2\right]_q e^{ikx} \end{eqnarray}
and
\begin{eqnarray}\int_{-\infty}^{\infty} dk\,\,(\tilde{\phi}_t(k,t) - \left[ -\nu k^2\right]_q \tilde{\phi}(k,t) ) e^{ikx} =0, \end{eqnarray}
which implies

$$\tilde{\phi}_t(k,t)=  \left[ -\nu k^2\right]_q \tilde{\phi}(k,t).$$

The general solution of the last equation is found in the form
\begin{eqnarray}
\tilde{\phi}(k,t)= \tilde{\phi}(k,0)\,\, e^{t \left[ -\nu k^2\right]_q }.\label{fourierresult}
\end{eqnarray}
Substituting (\ref{fourierresult}) into Fourier transform (\ref{fourier}) we get solution

$$\phi(x,t)= \frac{1}{\sqrt{2 \pi}} \int_{-\infty}^{\infty} dk\,\,e^{ikx+ t [-\nu k^2]_q } \tilde{\phi}(k,0). $$

By using the inverse Fourier transform, we can fix $\tilde{\phi}(k,0)$ by the initial function

$$\tilde{\phi}(k,0)=\frac{1}{\sqrt{2 \pi}}\int_{-\infty}^{\infty} dy\,\, \phi(y,0) e^{-iky}.  $$
Then solution of the initial value problem for infinite interval is

\begin{eqnarray}
\phi(x,t)= \int_{-\infty}^{\infty} G(x,y;t) \phi(y,0) dy\,,
\end{eqnarray}
where the Green function is defined as
\begin{eqnarray}
G(x,y;t)= \frac{1}{2 \pi} \int_{-\infty}^{\infty} e^{ik(x-y)+ t [-\nu k^2]_q} dk.
\end{eqnarray}

Using property (\ref{property1})

$$e^{t [-\nu k^2]_q} e^{ik (x-y)}= e^{t [\nu (ik)^2]_q }e^{ik(x-y)}=  e^{t [\nu \frac{d^2}{dx^2}]_q} e^{ik(x-y)},$$

the Green function becomes

$$G(x,y;t)= \frac{1}{2 \pi} \int_{-\infty}^{\infty} e^{t [\nu \frac{d^2}{dx^2}]_q} e^{ik(x-y)} dk = e^{t [\nu \frac{d^2}{dx^2}]_q} \frac{1}{2 \pi} \int_{-\infty}^{\infty} e^{ik(x-y)} dk. $$

As a result, the Green function for $q$-diffusive heat equation (\ref{infiniteinterval}) can be expressed as time evolution of the Dirac delta function
\begin{eqnarray}
G(x-y\,;t)= e^{t[\nu \frac{d^2}{dx^2}]_q} \delta(x-y).
\end{eqnarray}

\begin{definition}
The evolution operator is defined in terms of $q$-deformed operator as

\begin{eqnarray} U(t)= e^{t[\nu \frac{d^2}{dx^2}]_q}, \label{evolutionoperator}\end{eqnarray}
and gives evolution of the initial function $\phi(x,0)$
\begin{eqnarray}\phi(x,t)= e^{t[\nu \frac{d^2}{dx^2}]_q } \phi(x,0). \label{evolution} \end{eqnarray}
\end{definition}
As an example,
we consider the q-diffusive heat equation with initial value as the Dirac Delta function:
\begin{eqnarray}
\frac{\partial}{\partial t} \phi(x,t)&=& \left[\nu \frac{\partial^2}{\partial x^2}\right]_q \phi(x,t), \nonumber \\
\phi(x,0)&=& \delta(x).
\end{eqnarray}
 Then the solution is
\begin{eqnarray}
G(x,t)= e^{t [\nu \frac{d^2}{d x^2}]_q} \delta(x)= \frac{1}{2 \pi} \int_{-\infty}^\infty e^{ikx+ t[-\nu k^2]_q}\, dk.
\end{eqnarray}

\section{The Generalized Kampe-De Feriet Polynomials}
Here we are going to construct polynomial solutions of equation (\ref{qdeformedheatdeformation}).
For this we consider the plane wave solution of (\ref{qdeformedheatdeformation}) as the generating function for the Kampe de Feriet type polynomials,
$$\phi(x,t)= e^{kx+\omega(k) t},$$
which implies the q-deformed dispersion
$\omega(k)= [\nu k^2]_q$  and
\begin{eqnarray}
\phi(x,t)=e^{kx+ t[\nu k^2]_q}.\label{planewavesolution}
\end{eqnarray}
The phase velocity of this plane wave solution is characterized by $q:$  $$v_{ph}=\frac{\omega(k)}{k}= \frac{[\nu k^2]_q}{k}= \frac{e^{\nu k^2 \ln q}-1}{k(q-1)}.$$

\begin{figure}[!tbp]
  \centering
  \begin{minipage}[b]{0.45\textwidth}
    \includegraphics[width=\textwidth]{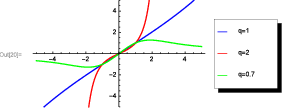}
    \caption{Phase Velocity}
    \label{phasevelocity}
  \end{minipage}
  \hfill
  \begin{minipage}[b]{0.45\textwidth}
    \includegraphics[width=\textwidth]{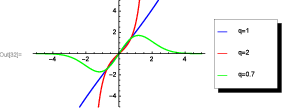}
    \caption{Group Velocity}
    \label{groupvelocity}
  \end{minipage}
\end{figure}

In Figure (\ref{phasevelocity}) we show the phase velocity for different values of $q$. In contrast to the linear dependence for $q=1$ case, for $q<1$ the velocity is bounded from above and reaches the maximum value and then it starts to decline fast. As we show in Section 5, this leads to a new process of soliton fissions.
However, for $q>1$ case the phase velocity is growing infinitely, that is, it has no upper limit.

The group velocity of this solution also depends on $q$ and is given by $$v_g=\frac{d \omega(k)}{d k}= \frac{2 \nu k \ln q}{q-1} e^{\nu k^2 \ln q}.$$

In Figure (\ref{groupvelocity}) we show  the group velocity for three different values of $q$. For $q < 1$ the group velocity is bounded from the above function,
taking maximal value
\begin{eqnarray}
|v_{max}| = \sqrt{\frac{2\nu}{e}\ln \frac{1}{q}}\frac{1}{1-q}
\end{eqnarray}
for $k = \pm 1/{\sqrt{2\nu \ln 1/q}}$.

\begin{definition}
The generalized Kampe-de Feriet polynomials $K_n(x,t)$ are defined as
\begin{eqnarray} e^{k x+ t [\nu k^2]_q}= \sum_{n=0}^\infty \frac{k^n}{n!} K_n(x,t). \label{KampedeFeriet}
\end{eqnarray}
\end{definition}

Property of these polynomials can be studied in a similar way as the usual Hermite and Kampe de Feriet polynomials. But in contrast to $q=1$ case, our generating function contains all powers of $k^2$  and requires introduction of the Bell polynomials. Before to proceed with this approach, in the next section we follow a more direct way by using dynamical symmetry of $q$-diffusive heat equation.

\section{Dynamical Symmetry for $q$-Diffusive Heat Equation}
For given differential equation $\hat{S} \phi=0$ with
\begin{eqnarray} \hat{S}= \frac{\partial}{\partial t}- H(P_1) \label{heat}\end{eqnarray}
exists the commuting operator $\hat{K}$  in the following form \cite{Pashaev2},
\begin{eqnarray}
\hat{K}= x+t H^{\,\,'}(P_1),\label{boost} \end{eqnarray}
where $P_1= \frac{d}{d x},$
such that $[\hat{S}, \hat{K}]=0.$
This $\hat{K}$ operator generates the dynamical symmetry for differential equation $\hat{S} \phi=0$: from given solution $\phi$ of the equation it creates another solution $\psi= \hat{K} \phi$ of the same equation $\hat{S} \psi=0$. The $\hat{K}$ operator in this form is linear in $x$ and $t$ and it represents the generalized Boost operator.

For our $q$-diffusive heat equation $$\frac{\partial}{\partial t} \phi= \left[\nu \frac{\partial^2}{\partial x^2}\right]_q \phi,$$
we have $H(P_1)= [\nu P_1^{\,\,2}]_q,$
and by taking derivative of $H(P_1)$ according to $P_1$
we get
$$H^{\,\,'}(P_1)= \frac{d}{d P_1} [\nu P_1^{\,\,2}]_q= \frac{d}{d P_1} \frac{e^{\nu \ln q P_1^{\,\,2}}-1 }{q-1}= \frac{2 \nu \ln q P_1}{q-1} e^{\nu \ln q P_1^{\,\,2}}.$$

%The Heat and the Boost operators for generic Hamiltonian $H$ we define as

%\begin{eqnarray}  \hat{S}&=& \frac{\partial}{\partial t}- H(P_1) \nonumber \\
% \hat{K}&=& x+t H^{\,\,'}(P_1). \label{boast} \end{eqnarray}

%From  (\ref{qdeformedheatdeformation}) by denoting $P_1= \frac{d}{d x},$ we have

Substituting the result into definitions  (\ref{heat}) and (\ref{boost}) we obtain the $q$-diffusive heat operator and the $q$-boost operator in the following form

\begin{eqnarray}
\hat{S}&=& \frac{\partial}{\partial t}- \left[ \nu \frac{\partial^2}{\partial x^2}\right]_q ,\label{qheat} \\
\hat{K}&=& x+ \frac{2 \nu \ln q}{q-1} t \frac{d}{d x} e^{\nu \ln q \frac{d^2}{d x^2}}. \label{qboost}
\end{eqnarray}

\begin{prop} The q-diffusive heat operator (\ref{qheat}) and the q-Boost operator (\ref{qboost})are commutative $[\hat{S},\hat{K}]=0.$
\end{prop}

\begin{prf}
\begin{eqnarray}
[S,K]&=& \left[\frac{\partial}{\partial t}-[\nu \frac{\partial^2}{\partial x^2}]_q,\, x\,+ \frac{2 \nu \ln q}{q-1} t \frac{d}{d x} e^{\nu \ln q \frac{\partial^2}{\partial x^2}} \right] \nonumber \\
&=& \frac{2 \nu \ln q}{q-1}  \frac{d}{d x} e^{\nu \ln q \frac{\partial^2}{\partial x^2}}\underbrace{\left[\frac{\partial}{\partial t},t\right]}_\text{1}-\underbrace{\left[[\nu \frac{\partial^2}{\partial x^2}]_q, x\right]}_\text{*}. \label{commut}
\end{eqnarray}

%\begin{eqnarray}
%[S,K]&=& \left[\frac{\partial}{\partial t}-[\nu \frac{\partial^2}{\partial x^2}]_q,\, x\,+ \frac{2 \nu \ln q}{q-1} t \frac{d}{d x} e^{\nu \ln q \frac{\partial^2}{\partial x^2}} \right] \nonumber \\
%&=& \frac{2 \nu \ln q}{q-1}  \frac{d}{d x} e^{\nu \ln q \frac{\partial^2}{\partial x^2}}\underbrace{\left[\frac{\partial}{\partial t},t\right]}_\text{1}-\underbrace{\left[[\nu \frac{\partial^2}{\partial x^2}]_q, x\right]}_\text{*}. \label{commutator}
%\end{eqnarray}

In order to find the commutator $(*),$  we use the following property:

%\begin{prop}
For any real analytic function $f(x)$ we have $$\left[f(\frac{d}{d x}),x\right]= f^{\,\,'}(\frac{d}{d x}),$$
%\end{prop}
%\begin{prf}
%If function $f(\frac{d}{d x})$ can be expanded to power series, we get the commutator
%$$[f(\frac{d}{d x}),x]=[\sum_{n=0}^\infty a_n \frac{d^n}{dx^n},x]= \sum_{n=1}^\infty a_n n \frac{d^{n-1}}{dx^{n-1}}=f^{'\,\,}(\frac{d}{d x}),$$
%where we have used $[\frac{d}{d x},x]=1$ and $[\frac{d^n}{d x^n},x]= n \frac{d^{n-1}}{dx^{n-1}}.$
%\end{prf}
which implies that the commutator $(*)$ can be written in the form:
\begin{equation}
\left[ \left[\nu \frac{\partial^2}{\partial x^2}\right]_q,\,x\right]=\left( [\nu \frac{\partial^2}{\partial x^2}]_q\right)^{'}.
\end{equation}

Calculating derivative of the operator
\begin{eqnarray}
\left( [\nu \frac{\partial^2}{\partial x^2}]_q\right)^{'}= \left( \frac{q^{\nu \frac{\partial^2}{\partial x^2}-1}}{q-1}\right)^{'},
\end{eqnarray}

and denoting $P\equiv \frac{\partial}{\partial x}$
we get
\begin{eqnarray}
\frac{d}{d P} \left(\frac{e^{\nu \ln q P^2}-1}{q-1} \right)= \frac{2 \nu P \ln q}{q-1}e^{\nu \ln q P^2}= \frac{2 \nu \ln q }{q-1}\frac{\partial}{\partial x}q^{\nu \frac{\partial^2}{\partial x^2}}.
\end{eqnarray}

Substituting the result into (\ref{commut}), finally we proved that $[\hat{S},\hat{K}]=0.$
\end{prf}

\begin{prop} If $\phi(x,t)$ is a solution of q-diffusive heat equation (\ref{qdeformedheatdeformation}) and $[\hat{S},\hat{K}]=0,$ then $\psi(x,t)= \hat{K} \phi(x,t)$ is also solution of this equation, where $\hat{S}$ is the $q$-dispersive Heat operator (\ref{qheat}) and $\hat{K}$ is the $q$-Boost operator (\ref{qboost}).
\end{prop}

According to this proposition: if $\phi(x,t)$ is a solution of the q-diffusive heat equation $\frac{\partial}{\partial t}\phi(x,t)= [\nu \frac{\partial^2}{\partial x^2}]_q \phi(x,t),$ then
\begin{eqnarray}\psi= \left( x+ t \frac{2 \nu \ln q}{q-1}\frac{\partial}{\partial x} e^{\nu \ln q \frac{\partial^2}{\partial x^2}}\right) \phi(x,t) \label{solutionof1} \end{eqnarray} is also solution.

\subsection{Bell Polynomials}

The generating function of Bell polynomials with $n$-variables is defined as
\begin{eqnarray}
\exp{\sum_{n=1}^\infty  \frac{g_n z^n}{n!}}= \sum_{n=0}^\infty B_n(g_1,q_2,...,g_n) \frac{z^n}{n!}, \label{generatingforbell}
\end{eqnarray}
and a few Bell polynomials are given below

\begin{eqnarray}
& & B_0=1, \nonumber \\
& & B_1(g_1)=g_1, \nonumber \\
& & B_2(g_1,g_2)= g_2+ g_1^2, \nonumber \\
& & B_3(g_1, g_2, g_3)= g_3 + 3g_1 g_2 + g_1^3. \nonumber
\end{eqnarray}
In particular case, when all independent variables are equal $g_1=g_2=...=g_n=x,$ the corresponding generating function (\ref{generatingforbell}) reduces to
the generating function for Bell polynomials of one variable $x$ defined in Wolfram MathWorld as
\begin{eqnarray} e^{x(e^z-1)}= \sum_{n=0}^\infty B_n(x) \frac{z^n}{n!}.\label{generatingbellwolf}\end{eqnarray}

A few Bell polynomials then are
\begin{eqnarray} B_0(x)=1,\,\,\, B_1(x)=x, \,\,\, B_2(x)=x+x^2,\,\,\, B_3(x)=x+3x^2+x^3,... \label{bell} \end{eqnarray}

\begin{prop}
The plane wave solution of  equation (\ref{qdeformedheatdeformation}) determines the  q-Kampe-de Feriet type polynomials $K_N(x,t;q)$
\begin{eqnarray} e^{kx} e^{[\nu k^2 ]_q t}= \sum_{N=0}^\infty \frac{k^N}{N!} K_N(x,t;q), \label{generatingkdf}
\end{eqnarray}
which can be represented in terms of the Bell polynomials $B_n(t)$:
\begin{eqnarray} K_N(x,t;q)= \sum_{n=0}^{[\frac{N}{2}]} \frac{x^{N-2n}N!}{(N-2n)! n!} B_n(\frac{t}{q-1}) (\nu \ln q)^n.\nonumber \end{eqnarray}
\end{prop}

\begin{prf}
By expanding the plane wave solution in $k,$ we have
\begin{eqnarray}
e^{k x} e^{[\nu k^2]_q t}=\left( \sum_{m=0}^\infty \frac{k^m}{m!} x^m\right) e^{[\nu k^2]_q t}.
\end{eqnarray}
Then, $e^{[\nu k^2]_q t}$ can be expanded in terms of Bell polynomials as
\begin{eqnarray}
e^{[\nu k^2]_q t}= e^{\frac{q^{\nu k^2}-1}{q-1}t}= e^{\frac{t}{q-1}(e^{\nu  \ln q k^2}-1)}= \sum_{n=0}^\infty B_n\left(\frac{t}{q-1}\right) \frac{(\nu  \ln q k^2)^n}{n!},
\end{eqnarray}
and the plane wave solution is written in the following form
\begin{eqnarray}
e^{k x} e^{[\nu k^2 ]_q t}= \sum_{m,n=0}^\infty \frac{k^{m+2n}}{m! n!} x^m B_n(\frac{t}{q-1}) (\nu \ln q )^n. \end{eqnarray}
By changing order of summation $m+2n=N,$

\begin{eqnarray}
e^{k x} e^{[\nu k^2 ]_q t}= \sum_{N=0}^\infty  \sum_{n=0}^{[\frac{N}{2}]}  \frac{k^N}{(N-2n)! n!} x^{N-2n} B_n \left(\frac{t}{q-1}\right)(\nu \ln q)^n \nonumber
\end{eqnarray}
we obtain the plane wave solution in the form of q-Kampe de Feriet type polynomials
\begin{eqnarray} e^{kx} e^{[\nu k^2 ]_q t}= \sum_{N=0}^\infty \frac{k^N}{N!} K_n(x,t;q),
\end{eqnarray}
where $$K_N(x,t;q)= \sum_{n=0}^{[\frac{N}{2}]} \frac{x^{N-2n}N!}{(N-2n)! n!} B_n\left(\frac{t}{q-1}\right) (\nu \ln q)^n.$$
\end{prf}

Using the first few Bell Polynomials (\ref{bell}) we can calculate $q$-Kampe De Feriet Polynomials
\begin{eqnarray}
 K_0(x,t;q)&=& 1  \nonumber \\
 K_1(x,t;q)&=& x \nonumber \\
 K_2(x,t;q)&=& x^2+\frac{2t}{q-1} \nu_q \nonumber \\
 K_3(x,t;q)&=& x^3+ \frac{6 t \nu_q }{q-1} \nonumber \\
 K_4(x,t;q)&=& x^4+12  t \frac{\nu_q}{q-1} x^2+12 (\frac{\nu_q t}{q-1})^2 +12 \frac{t}{q-1} (\nu_q)^2, \nonumber \\
& .&  \nonumber \\
& .&  \nonumber \\
& .&   \nonumber  \end{eqnarray}
where $\nu_q \equiv \nu \ln q.$
In the limit $q\rightarrow 1,$ these polynomials reduce to the standard Kampe de Feriet polynomials.

We can find the time evolution of zeros for these polynomials. For $n=2,$ we have two zeros evolving as
$$x_{1,2}=\pm \sqrt{\frac{2 t \nu_q}{1-q}}.$$
\begin{figure}[!tbp]
  \centering
  \begin{minipage}[b]{0.45\textwidth}
    \includegraphics[width=\textwidth]{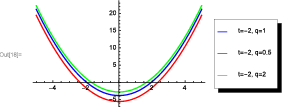}
    \caption{Motion of zeros at $t=-2$}
    \label{zeros2}
  \end{minipage}
  \hfill
  \begin{minipage}[b]{0.45\textwidth}
    \includegraphics[width=\textwidth]{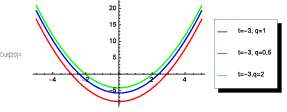}
    \caption{Motion of zeros at $t=-3$}
    \label{zeros3}
  \end{minipage}
\end{figure}

In Figures (\ref{zeros2}) and (\ref{zeros3}) we show the evolution of zeros, depending of values of $q.$ For $q<1,$ zeros are moving faster than $q=1$ case, and for $q>1,$ the motion slow down.

 In order to find the general form of these Kampe de Feriet polynomials for arbitrary $n$, we apply relation (\ref{solutionof1}) and
the Boost operator (\ref{qboost}).   Starting from $K_0(x,t;q)=1$ by successive application of this formula  we obtain
\begin{eqnarray}
 K_n(x,t;q)= \left( x+ \frac{2 t \nu_q}{q-1} \frac{\partial }{\partial x}e^{\nu_q \frac{\partial^2}{\partial x^2}}\right)^n \cdot 1.
\end{eqnarray}
As easy to see the polynomials result from evolution in time of monomials $$K_n(x,0;q)=x^n$$
by application of  evolution operator (\ref{evolution})
\begin{eqnarray}
K_n(x,t;q)= e^{t[\nu \frac{d^2}{dx^2}]_q} x^n.
\end{eqnarray}

\subsection{Dynamical Symmetry and Generating Function}
Here we like to find the generating function for our q-Kampe de Feriet polynomials (\ref{KampedeFeriet}) by application of the Boost operator

\begin{eqnarray}
 \sum_{n=0}^\infty \frac{k^n}{n!} K_n(x,t;q)= \sum_{n=0}^\infty \frac{k^n}{n!} \left( x+ \frac{2 t \nu_q}{q-1} \frac{d}{d x}e^{\nu_q \frac{d^2}{d x^2}}\right)^n \cdot 1 = e^{k(x+ \frac{2 t \nu_q}{q-1} \frac{d }{d x}e^{\nu_q \frac{d^2}{d x^2}})} \cdot 1
\end{eqnarray}
and show that it gives the plane wave solution (\ref{planewavesolution}).

\begin{prop} We have the following factorization formula
\begin{eqnarray}
e^{k(x+ \frac{2 t \nu_q}{q-1} \frac{d }{d x}e^{\nu_q \frac{d^2}{d x^2}})} \cdot 1= e^{k x} \,\, e^{[\nu k^2]_q t} \label{wavefactorize}
\end{eqnarray}
\end{prop}
To show this we need to use the Zassenhaus formula.

\begin{prop}
The Zassenhaus formula \cite{Magnus} for two operators $X$ and $Y$ is given by
\begin{equation}
e^{\xi(X+Y)}= e^{\xi X} e^{\xi Y} e^{-\frac{\xi^2}{2}[X,Y]} e^{\frac{\xi^3}{6}(2[Y,[X,Y]]+ [X,[X,Y]])}\,\,\,\, \cdot \cdot \cdot
\end{equation}
where $\xi$ is an arbitrary constant parameter.
\end{prop}

In order to apply the Zassenhaus formula to our case,
we denote $P\equiv\frac{d}{d x}, $
which satisfies the following commutation relations:
$$[P,x]=1,\,\,\,\, [P^2,x]=2P, \, \cdot \cdot \cdot [P^n,x]=n P^{n-1}$$ and can be generalized in the following form
\begin{eqnarray}
[f(P),x]= \frac{d}{dP}f(P).\label{generalizedcom}
\end{eqnarray}
In our formula (\ref{wavefactorize})  by changing variables

$$Y \equiv \frac{2t}{q-1} \nu_q \frac{d}{d x} e^{\nu_q \frac{d^2}{d x^2}}, \,\,\,\,\,\,\, Z \equiv e^{\nu_q \frac{d^2}{d x^2}},\,\,\,\,\, X\equiv x$$
 we obtain

\begin{eqnarray}
e^{k(x+ \frac{2 t \nu_q}{q-1} \frac{\partial }{\partial x}e^{\nu_q \frac{d^2}{d x^2}})} \cdot 1= e^{k(X+Y)}\cdot 1.
\end{eqnarray}
In order to factorize the exponential function we need to calculate the commutator relations:
$$[X,Y],\,\,\,\, [Y,[X,Y]],\,\,\,\, [X,[X,Y]],...$$

As easy to see all commutators are vanishing
\begin{eqnarray}
[[X,Y],Y]= [[[X,Y],Y],Y]=...=0
\end{eqnarray}
and therefore  the following commutator is zero
\begin{eqnarray}[P,Z]= [P, e^{e^{\nu_q P^2}}]=0.\nonumber \end{eqnarray}
Explicit calculation of $Z$ and $X$ commutator  is obtained in terms of the commutators of $X$ and $Y,$ which we need for the Zassenhaus formula

$$[Z,X] = [ e^{ \nu_q P^2}, X]= 2 \nu_q P e^{ \nu_q P^2}= \frac{q-1}{t} Y, $$
$$[[Z,X],X] =\frac{q-1}{t}[Y,X]= -\frac{q-1}{t}[X,Y], $$
$$[[[Z,X],X],X]= \frac{q-1}{t}[[Y,X],X]= (-1)^2 \frac{q-1}{t}[X,[X,Y]],$$
$$.$$
$$.$$
$$.$$
\begin{eqnarray}[[Z,\underbrace{X],X,...,X}_\text{n-times X}]= \frac{q-1}{t}[[Y,\underbrace{X],X,...,X}_\text{n-1 -times X}]= \frac{q-1}{t}[\underbrace{ X,[X,...,[X}_\text{n-1-times X},Y]. \label{commutator}\end{eqnarray}

Now let us find commutator of operators $Z$ and $X.$

Calculation of the following commutators give us derivatives
$$[Z,X]= [e^{\nu_q P^2},X]= \frac{d}{dP} e^{\nu_q P^2}= 2P\nu_q e^{\nu_q P^2}= 2 P\nu_q Z=\frac{d}{dP}Z$$
$$[[Z,X],X]=[2 P \nu_q Z, X]= 2 \nu_q (P[Z,X]+[P,X]Z)= 2 \nu_q (P \frac{d}{dP}+1)Z= 2 \nu_q (2 P^2 \nu_q +1)Z= \frac{d^2}{dP^2}Z$$
\begin{eqnarray}[[[Z,X],X],X]&=& 2 \nu_q ([2P^2 \nu_q Z+Z,X])= 2 \nu_q (2 \nu_q [P^2 Z,X]+ [Z,X]) \nonumber \\ &=& 2 \nu_q (2 \nu_q (P^2[Z,X]+ [P^2,X]Z)+ [Z,X]) \nonumber \\ &=& 2 \nu_q (4 \nu_q^2 P^3 Z+ 4 \nu_q PZ+ 2 \nu_q PZ)= \frac{d^3}{dP^3}Z, \end{eqnarray}
which can be generalized in the following proposition:
\begin{prop}

\begin{eqnarray}
[[Z,\underbrace{X],X,...,X}_\text{n-times X}]= \frac{d^n}{dP^n} Z= \frac{d^n}{dP^n} e^{\nu_q P^2} \label{commutation}
\end{eqnarray}
\end{prop}

It is easy to prove by mathematical induction.
%\begin{prf}
%From (\ref{generalizedcom}) $f(P)=Z(P)$
%\begin{eqnarray} & &[Z(P),X]= \frac{d}{dP} Z(P) \nonumber \\
%& &[[Z(P),X],X]= [\frac{d}{dP}Z(P),X]= \frac{d^2}{dP^2} Z(P) \nonumber \\
%& & [[Z(P),\underbrace{X],X,...X}_\text{n-times}]= \frac{d^n}{dP^n} Z(P)
%\end{eqnarray}
%\end{prf}

%By using mathematical induction
%we know that it is true: $$[Z,X]=  \frac{d}{d P} Z,$$ and let ts suppose it is true for $n:$
%\begin{equation}
%[[Z,\underbrace{X],X,...X}_\text{n-times X}]=\frac{d^n}{dP^n}Z
%\end{equation}
%and we should prove it for $n+1:$
%\begin{eqnarray}[[Z,\underbrace{X],X,...,X}_\text{n+1-times X}]= [\frac{d^n}{dP^n}Z, X]= \frac{d^n}{dP^n}[Z,X]+[\frac{d^n}{dP^n},X]Z=\frac{d^n}{dP^n} 2 &\nu_q P Z+ \frac{d}{dP}(\frac{d^n}{dP^n})Z \label{commutationrelation} \end{eqnarray}
%\begin{eqnarray}
%\frac{d^{n+1}}{dP^{n+1}}Z=\frac{d^{n+1}}{dP^{n+1}}(e^{\nu_q P^2})&=& \frac{d}{dP}(\frac{d^n}{dP^n} e^{\nu_q P^2})= \frac{d^{n+1}}{dP^{n+1}} e^{\nu_q P^2}+ %\frac{d^n}{dP^n}\frac{d}{dP}(e^{\nu_q P^2})\nonumber \\
%&=& (\frac{d^{n+1}}{dP^{n+1}}+ \frac{d^n}{dP^n}2 \nu_q P)e^{\nu_q P^2}= 2 \nu_q \frac{d^n}{dP^n} P Z+ \frac{d^{n+1}}{dP^{n+1}}Z
%\end{eqnarray}
%then we get the desired result in \ref{commutationrelation}
%$$[[Z,\underbrace{X],X,...,X}_\text{n+1-times X}]=2 \nu_q \frac{d^n}{dP^n} P Z+ \frac{d^{n+1}}{dP^{n+1}}Z =\frac{d^{n+1}}{dP^{n+1}}Z $$

\begin{prop}

The commutation relation (\ref{commutation}) can be expressed in terms of Hermite polynomials with operator argument
\begin{eqnarray}
[[Z,\underbrace{X],X,...X}_\text{n-times X}]=(-i)^n (\nu_q)^{\frac{n}{2}} H_n\left(i \sqrt{\nu_q} \frac{d}{dx}\right)e^{\nu_q \frac{d^2}{d x^2}}.
\end{eqnarray}
\end{prop}
\begin{prf}
From definition of Hermite polynomials
\begin{eqnarray}
H_n(\xi)= (-1)^n e^{\xi^2} \frac{d^n}{d \xi^n} e^{-\xi^2},
\end{eqnarray}
we have

\begin{eqnarray} H_n(\xi)e^{-\xi^2} =(-1)^n\frac{d^n}{d \xi^n} e^{-\xi^2}.
\end{eqnarray}
By considering commutation relation (\ref{commutation})
$$[[Z,\underbrace{X],X,...,X}_\text{n-times}]= \frac{d^n}{dP^n} Z= \frac{d^n}{dP^n} e^{\nu_q P^2},$$
and by changing variables
$\nu_q P^2 \equiv -\xi^2 \Rightarrow \xi=i \sqrt{\nu_q}P \Rightarrow \frac{dP}{d \xi}= \frac{-i}{\sqrt{\nu_q}},$

\begin{eqnarray}
H_n(\xi)e^{-\xi^2}&=& (-1)^n \frac{d^n}{d \xi^n} e^{-\xi^2} \nonumber \\
H_n(i \sqrt{\nu_q}P) e^{\nu_q P^2}&=& (-1)^n ( \frac{d P}{d\xi} \frac{d}{d P}  )^n e^{\nu_q P^2}= \frac{i^n}{(\nu_q)^{\frac{n}{2}}} \frac{d^n}{dP^n} e^{\nu_q P^2}
\end{eqnarray}

$$(-i)^n (\nu_q)^{\frac{n}{2}} H_n(i \sqrt{\nu_q}P)e^{\nu_q P^2}= \frac{d^n}{dP^n}e^{\nu_q P^2},$$
we can express commutation relation in terms of Hermite polynomials of the operator argument

\begin{eqnarray}
[[Z,X],X,...X]&=& \frac{d^n}{dP^n}e^{\nu_q P^2}= (-i)^n (\nu_q)^{\frac{n}{2}} H_n(i \sqrt{\nu_q}P)e^{\nu_q P^2} \nonumber \\
&=& (-i)^n (\nu_q)^{\frac{n}{2}} H_n(i \sqrt{\nu_q} \frac{d}{dx})e^{\nu_q \frac{d^2}{d x^2}}.
\end{eqnarray}
\end{prf}
Using the Zassenhaus formula
\begin{eqnarray}
e^{\xi(X+Y)}\cdot 1= e^{\xi X} e^{\xi Y} e^{-\frac{\xi^2}{2}[X,Y]} e^{\frac{\xi^3}{6}(2[Y,[X,Y]]+ [X,[X,Y]])} ...
e^{(-1)^{n+1}\frac{\xi^n}{n!}[X,[X,...,[X,Y]]]}...\cdot 1
\end{eqnarray}
%$$Y\equiv 2 \frac{t}{q-1}\nu_q \frac{d}{dx} e^{\nu_q \frac{d^2}{dx^2}},\,\,\,\,\, Z\equiv e^{\nu_q}\frac{d^2}{dx^2}$$

%$$[Z,X]= 2 \nu_q \frac{d}{dX}Z= 2 \nu_q \frac{d}{dx}e^{\nu_q \frac{d^2}{dx^2}}= \frac{q-1}{t}Y$$
%$$[[Z,X],X]= \frac{q-1}{t} [Y,X]$$
%$$[[[Z,X],X],X]= \frac{q-1}{t} [[Y,X],X]$$
%$$[[[Z,\underbrace{X],X],...,X}_\text{n times}]=  \frac{q-1}{t} [[[Y,\underbrace{X],X],...X}_\text{n-1 times}]$$
%$$[[[Y,\underbrace{X],X],...X}_\text{n-1 times}]= \frac{t}{q-1}[[[Z,\underbrace{X],X],...,X}_\text{n times}] $$
%and we know that $$[X,Y]=-[Y,X]$$
%$$[X,[X,Y]]=-[X,[Y,X]]= (-1)^2[[Y,X],X]$$
%\begin{eqnarray}[\underbrace{X,...,[X}_\text{n-1 times},Y]]= (-1)^{n-1}[[Y,\underbrace{X],...X}_\text{n-1 times}]&=&(-1)^{n-1} %\frac{t}{q-1}[[Z,\underbrace{X],X,...X}_\text{n times}] \nonumber \\
%&=& (-1)^{n-1} \frac{t}{q-1} (-i)^n (\nu_q)^{\frac{n}{2}} H_n(i\sqrt{\nu_q} \frac{d}{d x}) e^{\nu_q \frac{\partial^2}{\partial x^2}}
%\end{eqnarray}

and (\ref{commutator}) we can factorize the following exponential function as

\begin{eqnarray}
e^{\xi(x+2t\frac{\nu_q}{q-1}\frac{d}{d x} e^{\nu_q \frac{d^2}{dx^2}})} \cdot 1 &=& e^{\xi x}\, e^{\xi 2t \frac{\nu_q}{q-1}\frac{d}{dx}e^{\nu_q \frac{d^2}{dx^2}}} e^{\sum_{n=2}^\infty (-1)^{n+1} \frac{\xi^n}{n!}[X,[X,...,[X,Y]]]} \cdot 1 \nonumber \\
&=& e^{\xi x}\, e^{\xi 2t \frac{\nu_q}{q-1}\frac{d}{dx}e^{\nu_q \frac{d^2}{dx^2}}} e^{\sum_{n=2}^\infty \frac{\xi^n}{n!} \frac{t}{q-1} (-i)^n \nu_q^{\frac{n}{2}} H_n(i\sqrt{ \nu_q} \frac{d}{dx}) e^{\nu_q \frac{d^2}{dx^2}}} \cdot 1 \nonumber \\
&=& e^{\xi x}\, e^{\xi 2t \frac{\nu_q}{q-1}\frac{d}{dx}e^{\nu_q \frac{d^2}{dx^2}}}\prod_{n=2}^\infty e^{\frac{\xi^n}{n!} \frac{t}{q-1} (-i)^n \nu_q^{\frac{n}{2}} H_n(i \sqrt{\nu_q} \frac{d}{dx})e^{\nu_q \frac{d^2}{dx^2}}} \cdot 1 \nonumber \\
&=& e^{\xi x} \prod_{n=2}^\infty e^{\frac{\xi^n}{n!} \frac{t}{q-1} (-i)^n \nu_q^{\frac{n}{2}}H_n(0)}.
\end{eqnarray}

Due to relations for Hermite polynomials
\begin{eqnarray}
H_{2n}(0)&=&(-1)^n \frac{(2n)!}{n!}, \nonumber \\
H_{2n+1}(0)&=& 0 , \nonumber
\end{eqnarray}
we find that only terms with even numbers survive
\begin{eqnarray}
= e^{\xi x} \prod_{k=1}^\infty e^{ \frac{\xi^{2k}}{k!} \frac{t}{q-1} \nu_q^k }.
\end{eqnarray}

Replacing $\xi$ by $k$ and using $\nu_q=\nu \ln q$ we obtain

\begin{eqnarray} e^{k(x+2t \frac{\nu_q}{q-1} \frac{d}{dx} e^{\nu_q} \frac{d^2}{dx^2}) } \cdot 1= e^{kx} e^{\frac{t}{q-1} \sum_{l=1}^\infty \frac{k^{2l}}{l!} \nu^l (\ln q)^l }.
\end{eqnarray}
Finally we can factorize this expression in the form of the plane wave solution

\begin{eqnarray}
e^{k(x+2t \frac{\nu_q}{q-1} \frac{d}{dx} e^{\nu_q} \frac{d^2}{dx^2})} \cdot 1= e^{kx}\,\, e^{[\nu k^2]_q t}.
\end{eqnarray}

\section{$q$-Viscous Burgers' Equation}
We can relate our $q$-diffusive heat equation with nonlinear $q$-viscous Burgers' equation.
By dividing equation (\ref{qdeformedheatdeformation}) with $\phi(x,t)$
we obtain

\begin{eqnarray}
(\ln \phi(x,t))_t= \frac{1}{\phi(x,t)} \left[\nu \frac{\partial^2}{\partial x^2}\right]_q \phi(x,t)
\end{eqnarray}
 and taking the $x$ derivative from both sides and denoting

\begin{eqnarray}(\ln \phi(x,t))_x= \frac{\phi_x}{\phi}\equiv u ,\label{colehopf} \end{eqnarray} we get

\begin{eqnarray}
 u_t= \left(\frac{1}{\phi}\left[\nu \frac{\partial^2}{\partial x^2}\right]_q  \phi(x,t)\right)_x \label{burger1}
\end{eqnarray}

 \begin{prop} \label{prop2}
 \begin{equation}
 \frac{1}{\phi} \left[\nu \frac{\partial^2}{\partial x^2}\right]_q \phi= \left[\nu (\frac{d}{d x}+u)^2\right]_q \cdot 1, \label{burger2}
 \end{equation}
 where $u= \frac{\phi_x}{\phi}$
 \end{prop}

 \begin{prf}
 Using definition of the $q$-operator number
 \begin{eqnarray}
 \frac{1}{\phi} \left[\nu \frac{\partial^2}{\partial x^2}\right]_q \phi &=& \frac{1}{\phi} \frac{q^{\nu \frac{\partial^2}{\partial x^2}-1}}{q-1} \phi
 = \frac{1}{\phi} \frac{1}{q-1} (e^{\nu \ln q \frac{\partial^2}{\partial x^2}}-1)\phi \nonumber \\ &=& \frac{1}{q-1} \frac{1}{\phi} \sum_{n=1}^\infty \frac{(\nu \ln q)^n}{n!} \frac{\partial^{2n}}{\partial x^{2n}} \phi
 \end{eqnarray}
 and denoting $\phi \equiv e^f,$ which implies $$f= \ln \phi, \,\,\,\, f_x=(\ln \phi)_x = \frac{\phi_x}{\phi}\equiv u,$$
 we find
 \begin{eqnarray} \frac{1}{\phi} \left[\nu \frac{\partial^2}{\partial x^2}\right]_q \phi &=& \frac{1}{q-1} \sum_{n=1}^\infty \frac{(\nu \ln q)^n}{n!} (\frac{d}{d x}+u)^{2n} \cdot 1  =\frac{1}{q-1}(e^{\nu \ln q(\frac{d}{d x}+u)^2}-1) \cdot 1 \nonumber \\ &=& \frac{q^{\nu (\frac{d}{d x}+u)^2}-1}{q-1} \cdot 1 = \left[\nu \left(\frac{d}{d x}+u\right)^2\right]_q \cdot 1
 \end{eqnarray}
 \end{prf}

Substituting (\ref{burger2}) into equation (\ref{burger1}) we obtain the q-viscous Burgers' equation as

 \begin{eqnarray} u_t= \left(\left[\nu (\frac{d}{d x}+u)^2\right]_q \cdot 1\right)_x. \label{qdeformedburgers}
 \end{eqnarray}

%By using the Green's function for $q$-viscous heat equation (\ref{qdeformedheatdeformation}) and Cole-Hopf type substitution (\ref{colehopf}) we can solve the initial value problem form equation (\ref{qdeformedburgers}) in terms of IVP for the $q-$ viscous heat equation.

%(Calculations will be added.)

By using solution of $q$-diffusive heat equation we can find the solution of $q$-viscous Burgers' equation.
As a first particular solution of $q$-diffusive heat equation (\ref{qdeformedheatdeformation})
we choose the traveling  plane wave solution $$\phi(x,t)= e^{k x + [\nu k^2]_q t}.$$
This plane wave is the generating function for Kampe de Feriet polynomials (\ref{generatingkdf}), being polynomial solution of q-diffusive equation. Moving zeros of Kampe de Feriet Polynomials then correspond to moving poles of q-viscous Burgers' equation (\ref{qdeformedburgers}).

By using the Cole-Hopf transformation, the plane wave solution gives the constant solution of the q-viscous Burgers' equation $$u(x,t)= \frac{\phi_x}{\phi}= k.$$
 By considering the superposition of two plane waves with different wave numbers $k_1,k_2,$

 \begin{eqnarray}
 \phi(x,t)= e^{k_1 x + [\nu k_1^2]_q t}+ e^{k_2 x + [\nu k_2^2]_q t},
 \end{eqnarray}
we get shock soliton solution in the following form

 \begin{eqnarray}
 u(x,t)=\frac{\phi_x}{\phi}=\frac{k_1 e^{k_1 x + [\nu k_1^2]_q t}+k_2e^{k_2 x + [\nu k_2^2]_q t}} {e^{k_1 x + [\nu k_1^2]_q t}+ e^{k_2 x + [\nu k_2^2]_q t}} . \label{shocksoliton}
 \end{eqnarray}

\begin{figure}[!tbp]
  \centering
  \begin{minipage}[b]{0.4\textwidth}
    \includegraphics[width=\textwidth]{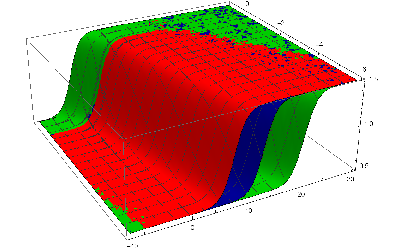}
    \caption{\small{One Shock Soliton for $q=1$(blue), $q=0.5$(red), $q=2$(green)}}
    \label{oneshock}
  \end{minipage}
  \end{figure}

In Figure (\ref{oneshock}) we show one shock soliton for different values of $q.$ Depending on value of $q$ the soliton is moving faster ($q<1$) or slower ($q>1$)
than in usual $q=1$ case. By fixing constants  $k_2>k_1 >0$, at fixed time we have assymptotics
\begin{eqnarray}
x \rightarrow + \infty &\Rightarrow& u \rightarrow k_2 \nonumber \\
x \rightarrow - \infty &\Rightarrow& u \rightarrow k_1 .\nonumber
\end{eqnarray}
Then our $q$-shock soliton solution can be written as

\begin{eqnarray}
u(x,t)=  \left( k_1+ \frac{k_2-k_1}{1+ e^{(k_2-k_1)(x-v t) }}\right),
\end{eqnarray}
where the velocity of shock is $$v=- \frac{[k_1^2 \nu]_q-[k_2^2 \nu]_q}{k_1-k_2}.$$
To analyze this expression we choose $k_1 = 0$ and denote $k_2 \equiv k$, so that the soliton velocity is
\begin{eqnarray} v= \frac{2 \nu k \ln q}{q-1} e^{\nu k^2 \ln q}.\end{eqnarray}
For $q < 1$ this velocity is bounded from the above, and takes
maximal value
\begin{eqnarray}
|v_{max}| = \sqrt{\frac{2\nu}{e}\ln \frac{1}{q}}\frac{1}{1-q}
\end{eqnarray}
for $k = \pm 1/{\sqrt{2\nu \ln 1/q}}$.

  \begin{figure}
  \centering
  \begin{minipage}[b]{0.4\textwidth}
    \includegraphics[width=\textwidth]{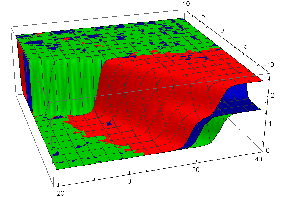}
    \caption{\small{Two Shock Solitons for  $q=1$(blue), $q=0.5$(red), $q=2$(green)}}
    \label{twoshock}
  \end{minipage}
\end{figure}

\begin{figure}[!tbp]
  \centering
  \begin{minipage}[b]{0.45\textwidth}
    \includegraphics[width=\textwidth]{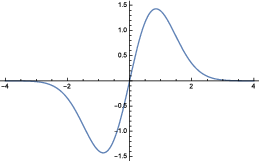}
    \caption{Soliton velocity for $q=0.5$}
    \label{solitonvelocity}
  \end{minipage}
  \hfill
  \begin{minipage}[b]{0.45\textwidth}
    \includegraphics[width=\textwidth]{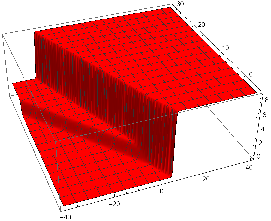}
    \caption{Soliton Fission for $q=0.5$}
    \label{solitonfission}
  \end{minipage}
\end{figure}

We show graph of this velocity in Figure (\ref{solitonvelocity}). This dependence creates a new property of the shock soliton.
Namely, for values of $k$ bigger than the extremum point, and corresponding amplitudes, the velocity is not growing, but
decaying. It produces new type of shock interaction. To see this we look for two shock soliton solutions.
By taking superposition of 3-plane waves
 \begin{eqnarray}
 \phi(x,t)= e^{k_1 x + [\nu k_1^2]_q t}+ e^{k_2 x + [\nu k_2^2]_q t}+e^{k_3 x + [\nu k_3^2]_q t} ,
 \end{eqnarray}
 we find two shock soliton solution in the form
 \begin{eqnarray}
 u(x,t)=\frac{\phi_x}{\phi}=\frac{k_1 e^{k_1 x + [\nu k_1^2]_q t}+k_2e^{k_2 x + [\nu k_2^2]_q t}+k_3e^{k_3 x + [\nu k_3^2]_q t}} {e^{k_1 x + [\nu k_1^2]_q t}+ e^{k_2 x + [\nu k_2^2]_q t}+ e^{k_3 x + [\nu k_3^2]_q t}}.
 \end{eqnarray}

In Figure (\ref{twoshock}) we show fusion of two shock solitons moving with speeds, depending on values of $q.$ For $q>1$ they move slower and for $q<1$
the speed of shocks collision is going faster than in $q=1$ case. In addition to this, for $q<1$ case here we have a new type of phenomena. By choosing
parameters $k_1 = 0$, $k_2 < k_0$ and $k_3 >k_0$, where $k_0 >0$ is extremum point with maximal speed, we find that the soliton with higher amplitude is moving slowly and splits to two solitons, one of which with smaller amplitude is moving faster. We illustrate this behavior in Figure (\ref{solitonfission}).

Superposition of $n+1$ plane waves with  wave numbers $k_1,k_2,...,k_{n+1}$ and constants $\eta_1,..., \eta_{k+1}$ gives $n$-shock soliton solution in the form

\begin{eqnarray}
u(x,t)= \frac{\sum_{i=1}^{n+1} k_i e^{k_i x + [\nu k_i^2]_q t+ \eta_i }}{\sum_{i=1}^{n+1} e^{k_i x + [\nu k_i^2]_q t+\eta_i }}.
\end{eqnarray}

\section{q - Semiclassical Expansion of $q$-diffusive heat equation}

If in $q$-diffusive heat equation (\ref{qdeformedheatdeformation}) we expand the right hand side according to $\nu,$
then we get infinite order equation with even order derivative in $x$,
\begin{eqnarray}
\frac{\partial \phi}{\partial t}=\left[\nu \frac{\partial^2}{\partial x^2}\right]_q \phi&=& \frac{q^{\nu \frac{\partial^2}{\partial x^2}-1}}{q-1} \phi \nonumber \\
&=& \frac{1}{q-1} (e^{\ln q \nu \frac{\partial^2}{\partial x^2}}-1)\phi \nonumber \\
&=& \frac{1}{q-1} \sum_{n=1^\infty} (\ln q)^n \nu^n (\frac{\partial^2}{\partial x^2})^n \phi \nonumber \\
&=& \frac{1}{q-1}(\nu \ln q \frac{\partial^2}{\partial x^2}+ \frac{(\nu \ln q)^2}{2!} \frac{\partial^4}{\partial x^4}+...)\phi, \label{deformationofheat}
\end{eqnarray}

where the first order equation for $\nu\ll 1,$ is the standard heat equation, but with deformed diffusion coefficient $\nu_q= \nu \frac{\ln q}{q-1}$,

\begin{eqnarray}\frac{\partial \phi}{\partial t}= \nu_q  \frac{\partial ^2}{\partial x^2} \phi. \label{1}\end{eqnarray}

In the limit $q\rightarrow 1$ case, this gives standard heat equation with diffusion coefficient $\nu.$

From another side, if we like to consider deformations of Heat equation for every power of $\ln q,$ which we called the q-semiclassical expansion (since $q=1$ case corresponds to "classical case") we need to use the Bernoulli polynomials.
\begin{prop}
Generating function for Bernoulli polynomials is defined as
\begin{eqnarray}
\frac{t e^{x t}}{e^t-1}= \sum_{n=0}^\infty B_n(x) \frac{t^n}{n!}. \label{generationforbernoulli}
\end{eqnarray}
For $x=0$ we have the generating function of Bernoulli numbers
\begin{eqnarray}
\frac{t}{e^t-1}= \sum_{n=0}^\infty B_n(0) \frac{t^n}{n!} \label{generatingforbernoullin0},
\end{eqnarray}
where $B_n(0)=B_n$-Bernoulli numbers.
\end{prop}

In the generating function of Bernoulli numbers (\ref{1}) by choosing $e^t\equiv q$ we have
\begin{eqnarray}
\frac{\ln q}{q-1}= \sum_{n=0}^\infty B_n \frac{(\ln q)^n}{n!}.
\end{eqnarray}

This gives the modified diffusion coefficient as expansion with Bernoulli numbers  in powers of $\ln q$
$$\nu_q=\nu \sum_{n=0}^\infty \frac{B_n}{n!} (\ln q)^n.$$

\begin{prop}
$[n]_q$ number can be  expressed in terms of Bernoulli polynomials as
\begin{eqnarray}
[n]_q= n+\sum_{m=1}^\infty \left( B_{m+1} (n)-B_{m+1}(0)\right) \frac{(\ln q)^m}{(m+1)!}
\end{eqnarray}
\end{prop}

\begin{prf}
Using the definition of $q$-numbers
\begin{eqnarray}[n]_q= \frac{q^n -1}{q-1}= \frac{e^{n \ln q}}{q-1}-\frac{1}{q-1}= \frac{1}{\ln q} \left( \frac{\ln q\,\, e^{n \ln q}}{q-1}\right)-\frac{1}{\ln q}\left( \frac{\ln q}{q-1}\right)
\end{eqnarray}
and denoting $\ln q= t,$
we obtain
\begin{eqnarray}
[n]_q=\frac{1}{t} \left( \frac{t e^{n t}} {e^t-1}\right)- \frac{1}{t} \left( \frac{t}{e^t-1}\right).
\end{eqnarray}

The generating function for Bernoulli polynomials (\ref{generationforbernoulli}) and (\ref{generatingforbernoullin0}) allow us to get
\begin{eqnarray}
[n]_q= \frac{1}{t}( \underbrace{B_0(n)-B_0(0)}_\text{* })+\sum_{m=0}^\infty \left( B_{m+1}(n)-B_{m+1}(0)\right) \frac{t^m}{(m+1)!},
\end{eqnarray}
$*=0$ due to $B_0(x)=1.$
And we can write
\begin{eqnarray}
[n]_q= \sum_{m=0}^\infty \left( B_{m+1}(n)-B_{m+1}(0)\right) \frac{t^m}{(m+1)!}= \underbrace{B_1(n)-B_1(0)}_\text{**}+ \sum_{m=1}^\infty \left( B_{m+1}(n)-B_{m+1}(0)\right) \frac{(\ln q)^m}{(m+1)!}, \nonumber
\end{eqnarray}
$**=n$ since $B_1(x)= x-\frac{1}{2}$ and the desired result is obtained.
\end{prf}

The $q$-number operator for an arbitrary operator $A$ can be written as a formal power series in terms of Bernoulli polynomials
\begin{eqnarray}
[A]_q= A+ \sum_{m=1}^\infty \left( B_{m+1} (A)-B_{m+1}(0)\right) \frac{(\ln q)^m}{(m+1)!}.
\end{eqnarray}

By expansion of $q$-diffusive heat equation (\ref{qdeformedheatdeformation}) in powers of $\ln q,$ we get higher derivative corrections to the Heat equation
\begin{eqnarray}
\frac{\partial \phi}{\partial t} = \nu \frac{\partial^2}{\partial x^2}\phi + \sum_{m=1}^\infty \left(B_{m+1} \left(\nu \frac{\partial^2}{\partial x^2}\right)-B_{m+1}\right) \frac{(\ln q)^m}{(m+1)!} \phi.
\end{eqnarray}

For the $q$-Galilean boost operator we obtain

\begin{eqnarray}
K= x+ 2 \nu t \frac{d}{d x}+ 2 \nu t \sum_{m=1}^\infty B_m\left(\nu \frac{d^2}{d x^2}\right) \frac{(\ln q)^m}{m!}.
\end{eqnarray}

The particular solution of $q$-diffusive heat equation for finite interval case can be expanded in the following form

\begin{eqnarray}
\phi_n (x,t)&=& e^{t[- \nu (\frac{\pi n}{l})^2]_q} \sin{\frac{n \pi}{l}x} \nonumber \\
&=& e^{-\nu (\frac{n \pi}{l}^2)x} \sin{\frac{n \pi}{l}x} \prod_{m=1}^\infty e^{(B_{m+1}(-\nu (\frac{n \pi}{t})^2 t) - B_{m+1}) \frac{(\ln q)^m}{(m+1)!}},
\end{eqnarray}
which shows how the solution of $q$-diffusive heat equation is modified by $q$-diffusitivity.

We can expand the Green function of $q$-diffusive heat equation for infinite interval case as

\begin{eqnarray}
G(x,y;t)&=& \frac{2}{l} \sum_{n=0}^\infty \sin{(\frac{n \pi}{l}x)} \sin{(\frac{n \pi}{l}y)}  e^{t [- \nu (\frac{n \pi}{l})^2]_q} \nonumber \\
&=&  \frac{2}{l} \sum_{n=0}^\infty \sin{(\frac{n \pi}{l}x)} \sin{(\frac{n \pi}{l}y)} e^{-\nu (\frac{n \pi}{l}^2)t}  \prod_{m=1}^\infty e^{(B_{m+1}(-\nu (\frac{n \pi}{t})^2 t) - B_{m+1}) \frac{(\ln q)^m}{(m+1)!}} \nonumber
\end{eqnarray}
showing modification due to  $q$-diffusivity.

The $q$-viscous Burgers' equation is also expandable in terms of higher order derivatives as an arguments of Bernoulli polynomials
\begin{eqnarray}
u_t&=& \left( \left[ \nu (\frac{d}{d x}+u)^2 \right]_q \cdot 1\right)_x \nonumber \\
&=& \nu u_{xx}+ 2 \nu u u_{x}+ \left(\sum_{m=1}^\infty (B_{m+1}\left(\nu \left(\frac{d}{dx}+u\right)^2\right)-B_{m+1}) \frac{(\ln q)^m}{(m+1)!} \cdot 1 \right)_x \nonumber
\end{eqnarray}

First two terms of this expansion give the standard Burgers' equation.

In a similar way we get expansion of the plane wave solution of $q$-diffusive heat equation in terms of powers of $\ln q,$

\begin{eqnarray}
\phi(x,t)= e^{k x+[\nu k^2]_q t}= e^{kx+\nu k^2 t }\prod_{m=1}^\infty e^{(B_{m+1}(\nu k^2 t)-B_{m+1}) \frac{(\ln q)^m}{(m+1)!}},
\end{eqnarray}
showing modification of the standard plane wave solution.

And by using the superposition of two travelling waves (\ref{shocksoliton}) with different wave numbers $k_1, k_2$ as a solution of $q$-diffusive heat equation
we obtain shock soliton solution of $q$-viscous Burgers' equation
as $q$- modification of standard shock soliton solution

\begin{eqnarray}
u(x,t)&=&\frac{\phi_x}{\phi}=\frac{k_1 e^{k_1 x + [\nu k_1^2]_q t}+k_2e^{k_2 x + [\nu k_2^2]_q t}} {e^{k_1 x + [\nu k_1^2]_q t}+ e^{k_2 x + [\nu k_2^2]_q t}} \nonumber \\
&=& \frac{k_1+ e^{(k_2-k_1)x + \nu (k_2^2-k_1^2)t} \prod_{m=1}^\infty e^{(B_{m+1}(\nu k_2^2 t)-B_{m+1}(\nu k_1^2 t)) \frac{(\ln q)^m}{(m+1)!}}}{1+e^{(k_2-k_1)x + \nu (k_2^2-k_1^2)t} \prod_{m=1}^\infty e^{(B_{m+1}(\nu k_2^2 t)-B_{m+1}(\nu k_1^2 t)) \frac{(\ln q)^m}{(m+1)!}} }
\nonumber \end{eqnarray}

For the speed of  $q$-shock soliton,
 $$v=- \frac{[k_1^2 \nu]_q-[k_2^2 \nu]_q}{k_1-k_2}$$
we have expansion
\begin{eqnarray}
v= -\left( \nu (k_1+k_2) + \sum_{m=1}^\infty \frac{B_{m+1}(k_1^2 \nu)- B_{m+1}(k_2^2 \nu)}{k_1-k_2} \frac{(\ln q)^m}{(m+1)!}\right).
\end{eqnarray}
By using explicit formula for Bernoulli polynomials for $n\geq 0$
\begin{eqnarray}
B_n(x)= \sum_{j=0}^n {n \choose j} b_j x^{n-j},
\end{eqnarray}
where $b_j$ are Bernoulli numbers,
the modified velocity is written as
\begin{eqnarray}
v= -\nu (k_1+ k_2) \left( 1 + \sum_{m=1}^\infty \frac{(\ln q)^m}{(m+1)!} S_{m+1} (k_1,k_2)\right),
\end{eqnarray}
where
\begin{eqnarray}
S_{m+1}\equiv \sum_{j=0}^{m+1} {m+1 \choose j} b_{m+1-j} [j]_{\nu k_1^2, \nu k_2^2}
\end{eqnarray}
and the $q$-number with $q_1,q_2$ basis are defined as

\begin{eqnarray}
[n]_{q_1,q_2}= \frac{q_1^n-q_2^n}{q_1-q_2}.
\end{eqnarray}

This shows the modification of the standard velocity of shock soliton.

Now we consider finite $\nu,$ but expand in terms of  $\epsilon=q-1 \ll 1 $ in (\ref{deformationofheat}).
Then for $q \rightarrow 1, \,\, \epsilon \ll 1,$  $q=1+\epsilon$ and $\nu$-arbitrary finite, we obtain higher order derivative corrections to heat equation

\begin{eqnarray}
\frac{\partial \phi}{\partial t}&=& \nu \frac{\partial^2}{\partial x^2} \phi+ \epsilon (-\frac{\nu}{2} \frac{\partial^2}{\partial x^2}+ \frac{\nu^2}{2} \frac{\partial^4}{\partial x^4})\phi + O(\epsilon^2) \nonumber \\
&=& \nu \left( 1-\frac{\epsilon}{2}\right) \frac{\partial^2}{\partial x^2} \phi+ \epsilon \frac{\nu^2}{2} \frac{\partial^4}{\partial x^4} \phi+ O(\epsilon^2),
\end{eqnarray}

In the first term, for small $q$ we have diffusion coefficient which is modified by $\epsilon.$

In the next section we are going to construct $q$-viscous Burgers' equation, related to our $q$-diffusive heat equation. For this we need next proposition:

\begin{prop}
\begin{eqnarray}
e^{-f} \frac{d^n}{d x^n} e^f= \left(\frac{d}{d x}+ f_x\right)^n \label{prop1}
\end{eqnarray}
\end{prop}

\begin{prf}
\begin{eqnarray}
e^{-f} \frac{d}{d x} e^f \psi= e^{-f} \frac{d}{d x}(e^f \psi)= e^{-f}(f_x e^f \psi+ e^f \frac{d}{d x}\psi)=(f_x+\frac{d}{d x})\psi,
\end{eqnarray}
so that
\begin{eqnarray}
e^{-f} \frac{d}{d x} e^f= f_x+ \frac{d}{d x}
\end{eqnarray}
and then we can generalize it as follows
\begin{eqnarray}
e^{-f} \frac{d^n}{dx^n} e^f&=& e^{-f} \frac{d}{d x} \frac{d}{d x}...\frac{d}{d x} e^f \nonumber \\
&=& e^{-f} \frac{d}{d x}e^f e^{-f} \frac{d}{d x}e^f e^{-f}...e^f e^{-f}\frac{d}{d x} e^f \nonumber \\
&=& \left(\frac{d}{d x}+f_x\right)^n.
\end{eqnarray}
\end{prf}

 \subsection{Corrections to $q$-viscous Burgers' Equation}
Expansion in $\epsilon= q-1$ provides higher derivative order corrections to the Burgers' equation.
 \begin{eqnarray}
 u_t&=& \left([\nu (\frac{d}{d x}+u)^2]_q\right)_x = \left(\frac{1}{q-1} (e^{\nu \ln q (\frac{d}{d x}+u)^2}-1)\cdot 1\right)_x= \left(\frac{1}{q-1} \sum_{n=1}^\infty \frac{(\nu \ln q)^n}{n!} (\frac{d}{dx}+u)^{2n} \cdot 1 \right)_x \nonumber \\
 &=& \frac{1}{q-1}\left(\frac{\nu \ln q}{1!} (\frac{d}{d x}+u)^2 \cdot 1 + \frac{(\nu \ln q)^2}{2!} (\frac{d}{d x}+u)^4 \cdot 1 +... \right)_x \nonumber \\
 &=& \frac{1}{q-1} \left(\nu \ln q \,\,(u_x+u^2)+ \frac{(\nu \ln q)^2}{2!} (u_{xxx}+ 4 u u_{xx}+3 u_x^2 + 6 u^2 u_x +u^4 )+... \right)_x \nonumber \\
 &=& \frac{\nu \ln q}{q-1} (u_{xx}+2 u u_x)+ \frac{(\nu \ln q)^2}{(q-1)2!} (...)+.... \label{correctionburger}
 \end{eqnarray}
 This gives the deformation of Burgers' equation with parameter $\nu.$
 In the limit $q\rightarrow 1$ it reduces to standard Burgers' equation
 \begin{eqnarray} u_t= \nu u_{xx}+2 \nu u u_{x}.\nonumber \end{eqnarray}

 For small $q=1+\epsilon,\,\,\, \epsilon \ll 1$ we write $$\ln q= \ln(1+\epsilon)= \epsilon -\frac{\epsilon^2}{2}+ \frac{\epsilon ^3}{3}-...= \epsilon(1-\frac{\epsilon}{2}+ \frac{\epsilon^2}{3}-...)$$ and after  substitution into (\ref{correctionburger}) we get

 \begin{eqnarray}
 u_t&=& \nu u_{xx}+ 2 \nu u u_x-\frac{\nu \epsilon}{2}(u_{xx}+2 u u_x)+ \frac{\nu^2}{2!} \epsilon (...)+O (\epsilon^2) \nonumber \\
 &=& (1-\frac{\epsilon}{2}) \nu u_{xx}+ 2(1-\frac{\epsilon}{2})  \nu u u_x+...
 \end{eqnarray}

This shows lower order corrections to Burgers equation from $q$-deformed viscosity.

 \section*{Acknowledgments}
This work was carried out with support from Izmir Institute of
Technology. One of the authors (S. Nalci T\"{u}mer) was partially supported by TUBITAK scholarship for graduate students.
O.K. Pashaev is supported by TUBITAK Research  Project TBAG: 116F206.

\end{document}